\documentclass[aps,reprint,superscriptaddress,nofootinbib,floatfix]{revtex4-1} 
\usepackage[english]{babel}

\usepackage{amsmath, amssymb}
\usepackage{mathtools}
\usepackage{siunitx}
\usepackage{environ}
\usepackage{mathrsfs}
\usepackage{braket}
\usepackage[colorlinks]{hyperref}
\DeclareMathAlphabet{\mathscrlower}{OT1}{pzc}{m}{it} 
\usepackage{prettyref}
\newrefformat{tab}{Table \ref{#1}}
\newrefformat{fig}{Figure \ref{#1}}
\usepackage{graphicx}
\usepackage{booktabs}
\usepackage{array}
\usepackage{multirow}
\usepackage{dcolumn}
\usepackage{threeparttable}
\usepackage{threeparttablex}
\usepackage{placeins}
\usepackage[stable]{footmisc}
\usepackage{mhchem}
\usepackage{soul}

\newcommand{\pauli}{\boldsymbol{\sigma}}
\newcommand{\Pauli}{\boldsymbol{\sigma}}
\newcommand{\diraccontra}[1]{\boldsymbol{\gamma}^{#1}}
\newcommand{\diracco}[1]{\boldsymbol{\gamma}^{#1}}

\newcommand{\diracg}{\vec{\boldsymbol{\gamma}}}

\newcommand{\diracs}{\vec{\boldsymbol{\Sigma}}}
\newcommand{\unity}{{\bf 1}_{2\times 2}}
\newcommand{\zeroty}{{\bf 0}_{2\times 2}}
\newcommand{\pos}{\vec{r}}
\newcommand{\mom}{\vec{p}}
\newcommand{\momop}{\hat{\vec{p}}}

\newcommand{\spinmom}{\vec{\Pauli}\cdot\momop}

\newcommand{\Sum}[2]{\sum\limits_{#1}^{#2}}

\newcommand{\parantheses}[1]{\left(#1\right)}
\newcommand{\brackets}[1]{\left[#1\right]}
\newcommand{\braces}[1]{\left\{ #1\right\}}
\let\nablatmp\nabla
\renewcommand{\nabla}{\vec{\nablatmp}}
\DeclarePairedDelimiter\abs{\lvert}{\rvert}
\makeatletter
\let\oldabs\abs
\def\abs{\@ifstar{\oldabs}{\oldabs*}}

\newcommand{\partiell}[2]{\frac{\partial #1}{\partial #2}}
\AtBeginDocument{
\heavyrulewidth=.08em
\lightrulewidth=.05em
\cmidrulewidth=.03em
\belowrulesep=.65ex
\belowbottomsep=0pt
\aboverulesep=.4ex
\abovetopsep=0pt
\cmidrulesep=\doublerulesep
\cmidrulekern=.5em
\defaultaddspace=.5em
}
\begin{document}
\title{Parity nonconserving interactions of electrons in chiral molecules with cosmic fields}
\date{\today}
\author{Konstantin Gaul}
\affiliation{Fachbereich Chemie, Philipps-Universit\"{a}t Marburg,
Hans-Meerwein-Stra\ss{}e 4, 35032 Marburg, Germany}

\author{Mikhail G. Kozlov}

\affiliation{Petersburg Nuclear Physics Institute of NRC
``Kurchatov Institute'', Gatchina 188300, Russia} %%
\affiliation{St.~Petersburg Electrotechnical University ``LETI'',
Prof. Popov Str. 5, 197376 St.~Petersburg}

\author{Timur A. Isaev}

\affiliation{Petersburg Nuclear Physics Institute of NRC
``Kurchatov Institute'', Gatchina 188300, Russia} %%

\author{Robert Berger}
\affiliation{Fachbereich Chemie, Philipps-Universit\"{a}t Marburg,
Hans-Meerwein-Stra\ss{}e 4, 35032 Marburg, Germany}
\begin{abstract}
Pseudoscalar or pseudovector cosmic fields, that serve as a source of
parity ($\mathcal{P}$) violation, are invoked in different models for
cold dark matter or in the standard model extension that allows for
Lorentz invariance violation. A direct detection of the
timelike-component of such fields requires a direct measurement of
$\mathcal{P}$-odd potentials or their evolution over time. Herein,
advantageous properties of chiral molecules, in which
$\mathcal{P}$-odd potentials lead to resonance frequency differences
between enantiomers, for direct detection of such $\mathcal{P}$-odd
cosmic fields are demonstrated. Scaling behavior of electronic
structure enhancements of such interactions with respect to nuclear
charge number and the fine-structure constant is derived analytically.
This allows a simple estimate of the effect sizes for arbitrary
molecules. The analytical derivation is supported by
quasi-relativistic numerical calculations in the molecules \ce{H2X2}
and \ce{H2XO} with X $=$ O, S, Se, Te, Po.  Parity violating effects
due to cosmic fields on the C--F stretching mode in CHBrClF are
compared to electroweak parity violation and influences of
non-separable anharmonic vibrational corrections are discussed. On
this basis it was estimated from a twenty year old experiment with
CHBrClF that bounds on Lorentz invariance violation as characterized
by the parameter $\abs{b^\mathrm{e}_0}$ can be pushed down to the order of
$10^{-17}\,\si{\giga\electronvolt}$ in modern experiments with
suitably selected molecular system, which will be an improvement of
the current best limits by at least two orders of magnitude. This
serves to highlight the particular opportunities that precision
spectroscopy of chiral molecules provides in the search for new
physics beyond the standard model. 
\end{abstract}
\maketitle
\section{Introduction}
In our recent work \cite{gaul:2020c} the virtues and prospects of chiral
molecules as direct sensors for pseudovector and pseudoscalar
cosmic fields were demonstrated. In the present paper we derive scaling laws
for interactions of electrons with these fields, presented in
\cite{gaul:2020c} and provide support from
numerical calculations. Furthermore, the methods applied for
derivation of limits on cosmic field interactions from experiments
with chiral molecules are presented in a more detailed manner and
accompanied by comparison to other computational methods.

One of the biggest puzzles of modern physics is the nature and composition
of dark matter (DM) (see e.g.
\cite{bertone:2005}). Many different models for dark matter exist,
considering objects that range from macroscopic to microscopic and
from being hot (ultra-relativistic) to cold (non-relativistic). Among
these DM theories cold DM (CDM)
theory serves to provide a simple explanation for many cosmological
observations~\cite{davis:1985}. However, the constituents of CDM are unknown
and can in principle fall in the range from macroscopic objects such as black holes to
new fundamental
particles like weakly interacting massive particles (WIMPs), 
axions, sterile neutrinos or dark photons (see e.g. Refs.
\cite{dodelson:1994,cheng:2002,arias:2012}). 

Despite its merits, the model of CDM has several
drawbacks~\cite{gentile:2004,klypin:1999,pawlowski:2014,kormendy:2010,sachdeva:2016,kroupa:2010}.
A possible solution of some of these provide fuzzy CDM models.  Fuzzy
CDM is supposed to consist of ultra light particles with masses of
$m_\phi\sim\SI{1e-22}{\electronvolt}/c^2$~\cite{hu:2000,lee:2018}.
This model makes searches for ultralight CDM oscillating with
frequencies on the order of \SI{1}{\micro\hertz} particularly
interesting.

CDM can consist of various types of weakly interacting
particles (an overview can be found e.g. in Ref. \cite{graham:2016}).
Among those pseudoscalar and pseudovector fields are of special interest as they are a
source of parity violation.

\emph{Pseudoscalar} CDM particles behave like axions, which were originally proposed
\cite{peccei:1977,wilczek:1978,weinberg:1978} to solve the strong $\mathcal{CP}$-problem of quantum
chromodynamics (QCD)~\cite{hooft:1976}. The search for CDM particles can
be restricted to
a comparatively small parameter space assessable to the QCD axion (see e.g.
\cite{cortona:2016}) or can involve a wide range for axionic particles
that are not bound to solve the strong $\mathcal{CP}$-problem. The
latter are often referred to as axion-like particles (ALPs).
\emph{Pseudovector} fields are important for models such as dark
photons~\cite{an:2015,catena:2018} and
also appear as sources of local Lorentz invariance violation in the  Standard
Model Extension (SME) by Kosteleck\'y and coworkers~\cite{colladay:1998}. 

In the last decade many new proposals for new experiments and improved
bounds on pseudoscalar CDM appeared, employing atomic spectroscopy (see e.g.
\cite{graham:2011,graham:2013,sikivie:2014,roberts:2014,stadnik:2014,graham:2018}).
Among those, strict limits on static $\mathcal{P}$-odd fields were set
from direct detection of parity violation with modern atomic precision
spectroscopy~\cite{roberts:2014,roberts:2014a}. In these experiments the dominating effect for parity
violation stems from the electroweak $Z^0$-mediated electron-nuclear
interaction.

Such $\mathcal{P}$-odd effects
are strongly enhanced in chiral molecules as well 
(for recent reviews on molecular parity violation
see~\cite{berger:2019,schwerdtfeger:2010,quack:2008,quack:2005,crassous:2005,berger:2004a,quack:2002}).
The chiral arrangement of the nuclei in the molecule leads to helicity of the electron cloud (see
e.g. Ref. \cite{berger:2004a}). Additional $\mathcal{P}$-odd effects
can then be measured as energy difference between enantiomers of chiral
molecules or as resonance frequency differences between the two
non-identical mirror-image molecules
\cite{quack:1986,gajzago:1974,letokhov:1975}. As
frequency shifts can be measured very accurately, this appears to be a
particularly promising tool to search for $\mathcal{P}$-odd cosmic fields.

In the following we analyse in detail the effects that emerge from
$\mathcal{P}$-odd cosmic fields in chiral molecules. We derive scaling
laws with respect to nuclear charge and the fine structure constant
and compare to what is known from parity violation due to electroweak
interactions. From our analysis we demonstrate advantages of the use
of chiral molecules to search for $\mathcal{P}$-odd cosmic fields. We
perform quasi-relativistic calculations at different levels of theory
and estimate the effect sizes in the vibrational spectra of the chiral
methane derivate CHBrClF~\cite{kompanets:1976,bauder:1997}. Thereby,
the computational difficulties are highlighted. From a twenty year old
experiment with this molecule~\cite{daussy:1999} we estimate the
sensitivity on cosmic parity violation \cite{gaul:2020c} and discuss
the scope for improvement on these limits in modern experiments with
chiral molecules and by improvement of present theoretical methods.

\section{Theory}
\subsection{Parity non-conserving interactions of electrons with cosmic
fields}
$\mathcal{P}$-odd interactions of electrons with pseudoscalar and pseudovector cosmic
fields were discussed in detail in Ref. \cite{roberts:2014a}.
A light pseudoscalar cosmic 
field obeys the Klein-Gordon equation. Assuming it to be
non-relativistic, i.e. $\hbar\omega_\phi\approx m_\phi c^2$ with $m_\phi$
being the CDM particle mass and $c$ being the speed of light in vacuum, we can write
\begin{equation}
\phi(\pos,t)=\phi_0\cos\left(\omega_\phi t - \frac{\pos\cdot\mom_\phi}{\hbar} +
\varphi\right),
\label{eq: cdmwave}
\end{equation}
where $\hbar=\frac{h}{2\pi}$ is the reduced Planck's constant,
$\phi_0$ is the CDM amplitude, $\mom_\phi=m_\phi\vec{v}_\phi$ is
the momentum of the CDM particle, which is proportional to its
velocity $\vec{v}_\phi$ and $\varphi$ is a phase factor. As the
relative velocity of the ALP field is suppressed by $10^{-3}$ with
respect to the speed of light (see Refs.
\cite{graham:2011,graham:2013} for details), for
terrestrial experiments we can assume
$\frac{\pos\cdot\mom_\phi}{\hbar}$ to be constant and choose
$\varphi$ such that eq. \prettyref{eq: cdmwave} can be written as
$\phi(\pos,t)=\phi_0\cos(\omega_\phi t)$ (see also Ref.
\cite{roberts:2014a}).

The interaction of the electronic field $\psi_\mathrm{e}$ with such 
pseudoscalar fields $\phi$ can be described by (see e.g.
\cite{wilczek:1978,weinberg:1978})
\begin{align}
 \label{eq: ac_pnc1}
 \mathcal{L}^\phi_{\mathrm{ps}}
 &=
 g_{\phi\mathrm{\bar{e}e}}(\hbar c\,\partial_\mu\phi)\bar{\psi}_\mathrm{e}\diraccontra{\mu}\diraccontra{5}\psi_\mathrm{e}\,,
\end{align}
where $g_{\phi\mathrm{\bar{e}e}}$ is a coupling constant of dimension
\si{\per\giga\electronvolt}. Herein the Dirac matrices are defined as 
\begin{equation}
\diraccontra{0}=\begin{pmatrix}
\unity &\zeroty\\
\zeroty &-\unity\\
\end{pmatrix},\qquad
\diraccontra{k}=\begin{pmatrix}
\zeroty &\pauli^k\\\
-\pauli^k &\zeroty\\
\end{pmatrix},
\end{equation}
where $\pauli^k$ are the Pauli spin matrices, $k=1,2,3$ and
$\mu=0,1,2,3$. 
$\diraccontra{5}=\imath\diraccontra{0}\diraccontra{1}\diraccontra{2}\diraccontra{3}$,
where $\imath=\sqrt{-1}$ is the imaginary unit,
$\partial_\mu=\partiell{}{x^\mu}$ is the first derivative with respect
to the four-vector $x^\mu=(ct, x, y, z)$ and Einstein's sum
convention is used.
Additionally a direct pseudoscalar coupling between the electrons and
the pseudoscalar cosmic field can be considered (see e.g.
Ref.~\cite{roberts:2014}):
\begin{align}
 \label{eq: ac_pnc2}
 \mathcal{L}^\phi_{\mathrm{dps}}
 &=-\imath
 \tilde{g}_{\phi\mathrm{\bar{e}e}} m_\mathrm{e} c^2 \phi\bar{\psi}_\mathrm{e}\diraccontra{5}\psi_\mathrm{e}\,,
\end{align}
where $\tilde{g}_{\phi\mathrm{\bar{e}e}}$ is a dimensionless coupling
constant and $m_\mathrm{e}$ is the mass of the electron. Whereas this
interaction can lead to parity violating couplings when considering
transition matrix elements of atomic or molecular
excitations~\cite{roberts:2014a}, it does not contribute to parity
violating expectation values, which give dominant contributions to
frequency differences in spectra of chiral molecules. Thus these
interactions are not discussed any further in the following.

The time-derivative of the pseudoscalar field leads
to the $\mathcal{P}$-odd single-electron Hamiltonian 
\begin{equation}
\hat{h}_\mathrm{ps}=g_{\phi\mathrm{\bar{e}e}}
\sqrt{2(hc)^3\rho_\mathrm{CDM}}
\sin(\omega_\phi t)
\diraccontra{5},
\label{eq: alppv}
\end{equation}
where
$\rho_\mathrm{CDM}\approx\frac{(\hbar\omega_\phi\phi_0)^2}{2(hc)^3}$
is the CDM energy density, for which we assume all ALPs  to comprise all of the
CDM with a uniform density:
$(hc)^{3}\rho_\mathrm{CDM}=(hc)^{3}\SI{0.4}{\giga\electronvolt\per\centi\meter\cubed}=\SI{7.6e-4}{\electronvolt^4}$
(see Ref. \cite{vergados:2017}). We use lowercase letters ($\hat{h}$) for 
single-electron operators and uppercase letters ($\hat{H}$) for multi-electron
operators. These are in the case of $\hat{H}_\mathrm{ps}$ (as well as
$\hat{H}_\mathrm{pv}$, $\hat{H}_\mathrm{ew}$ given below) simple sums over
all electrons of the system, e.g. $\hat{H}_\mathrm{ps}=\sum_i
\hat{h}_\mathrm{ps}(i)$

Electronic interactions with pseudovector cosmic fields
can be described by the Lagrangian 
\begin{equation}
\mathcal{L}_\mathrm{pv}^b = - b_\mu
\bar{\psi}_\mathrm{e}\diraccontra{\mu}\diraccontra{5}\psi_\mathrm{e},
\end{equation}
which appears e.g. in the local Lorentz invariance violating Standard Model
Extension (SME) (for details
see Refs. \cite{colladay:1998,kostelecky:1999}).

The 
parity non-conserving interaction Hamiltonian for the temporal
component is
\begin{equation}
\hat{h}_\mathrm{pv}=b_0(t)\diraccontra{5},
\label{eq: pseudovector}
\end{equation} 
where the field can be static $b_0(t)=b^\mathrm{e}_{0}$ or dynamic
$b^\mathrm{e}_0(t)=b^\mathrm{e}_0\sin(\omega_b t)$. Here
$b^\mathrm{e}_0$ is the interaction strength of the timelike-component of
the pseudovector field with the electrons.

In spectra of chiral molecules the interactions discussed above lead to shifts (static fields) or
oscillations (dynamic fields) of frequency shifts due to the nuclear
spin-independent electroweak interactions, the main
contribution to which is in closed-shell molecules expected to arise from 
the electron-nuclei weak neutral-current interaction Hamiltonian (see e.g.
\cite{berger:2019,schwerdtfeger:2010}):
\begin{equation}
\hat{h}_\mathrm{ew}=\frac{G_\mathrm{F}}{2\sqrt{2}}\Sum{A=1}{N_\mathrm{nuc}}Q_{\mathrm{W},A}\rho_A(\pos)\diraccontra{5}\,,
\label{eq: ewpvoperator}
\end{equation}
where $G_\mathrm{F}=\SI{2.22249e-14}{\hartree\bohr\cubed}$ is Fermi's
weak coupling constant, $Q_{\mathrm{W},A}$ and $\rho_A$ are the weak
charge and normalized charge density of nucleus $A$, respectively. The
total number of nuclei is $N_\mathrm{nuc}$. Contributions from
$\mathcal{P}$-odd nuclear-spin dependent terms when combined with
$\mathcal{P}$-even hyperfine coupling \cite{gorshkov:1982,hobi:2013}
are estimated to give only minor contributions in closed-shell
molecules. Similar considerations hold for the contribution from
neutral-current interaction terms between electrons. 

It shall be noted that in chiral molecules weakly interacting dark
matter candidates, such as WIMPs, or cosmic neutrinos can also lead to
shifts or oscillations of the $\mathcal{P}$-odd potential as was
discussed by Bargue\~no
\emph{et.al.}~\cite{bargueno:2006,bargueno:2007,bargueno:2008}. These
interactions as well as those of electrons with pseudoscalar and
pseudovector fields discussed above are proportional to
$\Braket{\diraccontra{5}}$.  In the following we will discuss in
general the chiral operator $\diraccontra{5}$, which leads to parity
non-conservation and compare to known properties of operator
\prettyref{eq: ewpvoperator}.

\subsection{Molecular expectation value of
$\diraccontra{5}$\label{molexpec}}
The time-independent Dirac-Coulomb equation for the electronic system of
the molecule reads
\begin{align}\label{Dirac1}
 \hat{H}_{\mathrm{DC}}\Psi_I = E_I \Psi_I,
\end{align}
with $\Psi_I$ and $E_I$ being the $I$th eigenfunction and eigenvalue
of the Dirac-Coulomb Hamiltonian being given by
\begin{equation}
\begin{split}
\hat{H}_{\mathrm{DC}}=&\Sum{i}{N_\mathrm{elec}}\left[c\diraccontra{0}\diracg\cdot\momop_i+\parantheses{\diraccontra{0}-\mathbf{1}}m_\mathrm{e}c^2
\phantom{\Sum{j\ne i}{N_\mathrm{elec}}}\right.\\
&\left. +V_\mathrm{nuc}(\pos_i)+\frac{1}{2}\Sum{j\ne i}{N_\mathrm{elec}}k_\mathrm{es}\frac{e^2}{\abs{\pos_i-\pos_j}}\right] \,,
\end{split}
\end{equation}
where we shifted the energy levels by $-m_\mathrm{e} c^2$ to bring the
upper part of the spectrum into correspondence with the
non-relativistic limit of the energy levels.  Here
$e$ is the elementary electric charge, $k_\mathrm{es}$ is in SI units $\frac{1}{4\pi\epsilon_0}$ with
$\epsilon_0$ being the electric constant and $V_\mathrm{nuc}$ being the
potential the nuclei in the molecule produce.

In the Dirac--Hartree--Fock--Coulomb (DHFC) approach, the multi-electron states $\Psi_I$ are
approximated by a Slater determinant build from an orthonormal set of
single-electron bi-spinors $\psi_i$ with orbital energy $\epsilon_i$. From the lower equation of
the resulting single-electron Dirac equations
expressions for the lower components $\chi_i$ of the 
Dirac bi-spinors 
\begin{equation}
 \psi_{i}(\pos) =
 \begin{pmatrix}
 \varphi_{i}(\pos)\\
 \chi_{i}(\pos)
 \end{pmatrix}
\end{equation}
can be found via
 \begin{equation} \label{eq: chi}
 \chi_{i}(\pos)
=c\parantheses{2m_\mathrm{e}c^2-\hat{V}+\varepsilon_i
}^{-1} \spinmom\,\varphi_{i}(\pos),
\end{equation}
where we have omitted all multi-electron effects for the sake of
simplifying the discussion below. 

For the remaining part of this section we will use atomic units, in which $\hbar$, $\abs{e}$ and 
$m_\mathrm{e}$ have the numerical value of 1. Then, the term in
parentheses in eq.\ \eqref{eq: chi} can be expanded in orders of the fine structure constant
$\alpha=c^{-1}$ as
\begin{equation}
c\parantheses{2c^2-\hat{V}+\varepsilon_i}^{-1}
=\frac{\alpha}{2}\Sum{k=0}{\infty}\brackets{\frac{\alpha}{2}\parantheses{\hat{V}-\varepsilon_i}}^k\,.
\end{equation}
Truncation after first order yields the Pauli approximation:
 \begin{equation}
 \chi_{i}(\pos)
=\brackets{\frac{\alpha}{2}+\frac{\alpha^3}{4}\parantheses{\hat{V}-\varepsilon_i}}
\spinmom\,\varphi_{i}(\pos).
\label{eq: pauli}
\end{equation}

In a molecule, the expectation value of $\diraccontra{5}$ for a single
Slater determinant is determined by a summation over contributions from all
occupied molecular orbitals $i$:
\begin{equation}
\Braket{\psi_i|\diraccontra{5}|\psi_i}=\Braket{\varphi_i|\chi_i}+\Braket{\chi_i|\varphi_i}
\label{eq: gamma50}
\end{equation}
Insertion of the first term of the expansion \prettyref{eq: pauli} in
eq. \prettyref{eq: gamma50} gives the first order contribution to
$\diraccontra{5}$:
\begin{equation}
\Braket{\psi_i|\diraccontra{5}|\psi_i}\approx\alpha\Braket{\varphi_i|\spinmom|\varphi_i}
\end{equation}
This obviously vanishes if the overall electron density of the
molecule is non-helical, but can, in the static case and when remaining in
first order with respect to $\mathcal{P}$-odd operators, only be non-zero for 
a chiral molecule, in which the electron density can have non-vanishing 
helicity.  

In order to determine scaling laws with respect to the nuclear charge
number $Z$ and the fine-structure constant $\alpha$, eq. \prettyref{eq:
gamma50} itself is not immediately useful. This is why we follow
Ref. \cite{roberts:2014} and write the operator $\diraccontra{5}$ 
for electron $i$ as a commutator:
\begin{align}\label{eq: gamma5}
 \diraccontra{5}_i
 &=
 \frac{\imath}{c}\left[\hat{H}_\mathrm{DC},\diracs_i\cdot\pos_i\right]_-
 +2
 \begin{pmatrix}
 \mathbf{0} & \hat{\mathbf{k}}_i\\
 \hat{\mathbf{k}}_i & \mathbf{0}
 \end{pmatrix},
 \\
 \label{hatK}
 \hat{\mathbf{k}}_i &= -(\mathbf{1}_i+\vec{\pauli}_i\cdot\hat{\vec{l}}_i)\,,
 \qquad
 \diracs_i = 
  \begin{pmatrix}
 \vec{\pauli}_i & \mathbf{0}\\
 \mathbf{0} & \vec{\pauli}_i
 \end{pmatrix}\,.
\end{align}
Eigenvalues of the operator
$\hat{\mathbf{K}}=\sum_i\hat{\mathbf{k}}_i$ in atomic systems
correspond to the  relativistic quantum numbers
$\varkappa=(\ell-j)(2j+1)$, where $\ell$ and $j$ are the orbital and
total angular momentum quantum numbers, respectively.

As long as we are interested in expectation values of the operator
$\diraccontra{5}$ on the molecular DHFC-orbitals $\psi_i$, the
commutator part in eq. \prettyref{eq: gamma5} turns to zero. DHFC molecular
orbital matrix elements of the second term in eq. \prettyref{eq: gamma5}
have the form
\begin{align}\label{eq: gamma5a}
\Braket{\psi_i|\diraccontra{5}|\psi_i}
=
 2\Braket{\varphi_i| \hat{\mathbf{k}}|\chi_i}
 +
 2\Braket{\chi_i| \hat{\mathbf{k}}|\varphi_i}.
\end{align}
The non-relativistic limit of $\Braket{\diraccontra{5}}$ vanishes as
can be shown by insertion of the first term of the expansion
\prettyref{eq: pauli} in eq. \prettyref{eq: gamma5a}:
\begin{align}\label{gamma5NR}
 \Braket{\psi_i| \diraccontra{5}|\psi_i}
 \approx
 \alpha\Braket{\varphi_i|\braces{\hat{\mathbf{k}},\spinmom}_+|\varphi_i}
 =0\,,
 \end{align}
where we use the fact that operator $\hat{\mathbf{k}}$ anti-commutes with
$\spinmom$:
\begin{equation}\label{eq: commut1}
\braces{\hat{\mathbf{k}},\spinmom}_+=0.
\end{equation}
The terms of order $\alpha^3$ give:
\begin{widetext}
\begin{align}\label{eq: gamma5rel}
 \Braket{\psi_i| \diraccontra{5}|\psi_i} 
 %\\
 \approx
 \frac{\alpha^3}{2}
 \Braket{\varphi_i|
 (\spinmom)\hat{V}\hat{\mathbf{k}}+\hat{\mathbf{k}}\hat{V}(\spinmom)
 |\varphi_i}
 \,,
 \end{align}
where the terms containing orbital energies $\varepsilon_i$
reduce to the anti-commutator \eqref{eq: commut1}.
Equation \eqref{eq: gamma5rel} can be rewritten as:
\begin{multline}\label{eq: mcommut}
\Braket{\!\psi_i|\diraccontra{5}|\psi_i}
\approx
\frac{\alpha^3}{2}\Braket{\varphi_i|
 \brackets{\spinmom,\hat{V}(\pos)}_-\hat{\mathbf{k}}
%|\varphi_i}
%\\
+
%\Braket{\varphi_i|
\hat{V}(\pos)\braces{\hat{\mathbf{k}},\spinmom}_+\!\!
+\brackets{\hat{\mathbf{k}},\hat{V}(\pos)}_-\spinmom
|\varphi_i\!}
\\
=
\frac{\alpha^3}{2}\Braket{\varphi_i|
 \brackets{\spinmom,\hat{V}(\pos)}_-\hat{\mathbf{k}}
+\brackets{\hat{\mathbf{k}},\hat{V}(\pos)}_-\spinmom
|\varphi_i\!},
\end{multline}
\end{widetext}
where we once again used eq.\ \eqref{eq: commut1}.
In general, the molecular potential energy operator $\hat{V}$ does not
commute with both operators $\hat{\mathbf{k}}$ and $(\spinmom)$.
However, its spherically symmetric part
$\hat{V}_\mathrm{s}(\abs{\pos})$ commutes with the operator
$\hat{\mathbf{k}}$. Therefore, for the spherically symmetric potential
the last term in eq. \eqref{eq: mcommut} turns to zero. Let us separate the
contribution of $\hat{V}_\mathrm{s}(\abs{\pos})$:
\begin{align}\label{eq: gamma_s-a}
\Braket{\psi_i|\diraccontra{5}|\psi_i}            &=
                    \Braket{\psi_i|\diraccontra{5}|\psi_i}_\mathrm{s}
                   +\Braket{\psi_i|\diraccontra{5}|\psi_i}_\mathrm{a}\,,
\\   \label{eq: gamma_s1}                
\Braket{\psi_i|\diraccontra{5}|\psi_i}_\mathrm{s} &= \frac{\alpha^3}{2}
                                      \Braket{\varphi_i|
                                       -\imath(\vec{\pauli}\cdot\pos) 
                                       \frac{\hat{V}'_\mathrm{s}(\abs{\pos})}{\abs{\pos}}\hat{\mathbf{k}} 
                                      |\varphi_i}\,
\end{align}
and consider the term \eqref{eq: gamma_s1} in more detail.  
Note that ${\hat{V}'_\mathrm{s}(\abs{\pos})}/{\abs{\pos}}$ commutes with both operators
$\vec{\pauli}\cdot\vec{r}$ and $\hat{\mathbf{k}}$. By analogy with
\prettyref{eq: commut1} we can assume that
$\braces{\vec{\pauli}\cdot\pos,\hat{\mathbf{k}}}_+=0$.
Thus, we can write:
 \begin{equation}\label{navarot}
 \imath(\vec{\pauli}\cdot\pos) \hat{\mathbf{k}}
 = \frac{\imath}{2}\brackets{\vec{\pauli}\cdot\pos,\hat{\mathbf{k}}}_-\,,
 \end{equation}
which proves that the operator in \eqref{eq: gamma_s1} is hermitian,
and allows to rewrite this expression as:
\begin{align}\label{eq: gamma_s2}
\Braket{\psi_i|\diraccontra{5}|\psi_i}_\mathrm{s} &= \frac{\alpha^3}{4}
                                        \Braket{\varphi_i|
                                        \vec{\pauli}\cdot\vec{v}_{\mathcal{T},\mathrm{s}}
                                        |\varphi_i}, 
\\ \label{eq: gamma_s3}
\vec{v}_{\mathcal{T},\mathrm{s}} &=
                                   \frac{\hat{V}'_\mathrm{s}(\abs{\pos})}{\abs{\pos}}
                                   \parantheses{\abs{\pos}^2\momop - \pos (\momop\cdot\pos)}.
\end{align}

We see that expectation value \eqref{eq: gamma_s2} has the form of a scalar product of the spin
with an electronic orbital $\mathcal{T}$-odd vector $\vec{v}_{\mathcal{T},\mathrm{s}}$.
Molecular matrix elements of $\vec{\pauli}\cdot\vec{v}_\mathcal{T}$ turn to zero in the
non-relativistic approximation for two reasons: (i) for a singlet
state an expectation value of the spin is zero; (ii) matrix elements
of orbital $\mathcal{T}$-odd vectors are imaginary, so their
expectation values are zero. In order to get a non-zero expectation
value of such operators one needs to include spin-orbit interactions
$\hat{H}_\mathrm{so}$, which mix singlet and triplet molecular states
and have imaginary matrix elements. Therefore, the energy shift
$\delta E_{\diraccontra{5},\mathrm{s}}$ of
the molecular (ground) singlet state due to the interaction
$\vec{\pauli}\cdot\vec{v}_{\mathcal{T},\mathrm{s}}$ appears in
double perturbation theory as:
 \begin{align}
 \label{Eshift}
 \delta E_{\diraccontra{5},\mathrm{s}}=
 \frac{\alpha^3}{2}
\frac{\mathfrak{Re}\braces{\Braket{\Psi_\mathrm{s}|\vec{\pauli}\cdot\vec{v}_{\mathcal{T},\mathrm{s}}
|\Psi_\mathrm{t}}\Braket{\Psi_\mathrm{t}|\hat{H}_\mathrm{so}|\Psi_\mathrm{s}}}}
 {E_\mathrm{s}-E_\mathrm{t}}
 \,.
 \end{align}
where $E_\mathrm{s}$, $E_\mathrm{t}$ and $\Psi_\mathrm{s}$,
$\Psi_\mathrm{t}$ are the non-relativistic singlet and triplet
energies and wave functions, respectively.

Equation \eqref{Eshift} allows to estimate the scaling law for $\delta
E_{\diraccontra{5},\mathrm{s}}$ with the nuclear charge $Z$ and the
fine structure constant $\alpha$. The matrix element of the spin-orbit
interaction
$\Braket{\psi_\mathrm{t}|\hat{H}_\mathrm{so}|\psi_\mathrm{s}}$ scales
as $\alpha^2 Z^2$. The $Z$ scaling of the matrix element of the
operator $\vec{v}_{\mathcal{T},\mathrm{s}}$ depends on the distances
where the integral is accumulated. Taking into account that this
operator appears in third order in $\alpha$, we can assume that the
integral is accumulated at short distances near the nucleus, where
relativistic corrections are larger. At such distances the potential
of the nucleus is practically unscreened, $\hat{V}_\mathrm{s}\sim
Z/r$. Furthermore, at these distances the electron moves $Z$ times
faster, so $\momop\sim Z$. Therefore, we can assume that  $\int
{v}_{\mathcal{T},\mathrm{s}} d^3r \sim Z^2$. Then the overall scaling
is:
 \begin{align}
 \label{E_s_scaling}
 \delta E_{\diraccontra{5},\mathrm{s}}\sim \alpha^5 Z^{4} \,.
 \end{align}

The last expression does not take into account ``the single center
theorem'' \cite{hegstrom:1980,kozlov:1982},
which implies that electron helicity in molecules is suppressed in the
vicinity of a single heavy nucleus and one 
has to take two matrix elements of expression \eqref{Eshift} at two different 
heavy centers. Therefore, the final scaling should be:  
 \begin{align}
 \label{E_s_scaling1}
 \delta E_{\diraccontra{5},\mathrm{s}}\sim \alpha^5 Z_A^{2}Z_B^2\,,
 \end{align}
where $A$ and $B$ are typically taken as the two heaviest atoms in the molecule.

Now let us analyze the second term in eq.\ \eqref{eq: gamma_s-a}. In
this case both terms from eq.\ \eqref{eq: mcommut} can contribute. For
the first term we can use the same arguments as above, but the
asymmetric part of the molecular potential at short distances is much
weaker, so this term will add small corrections to eq.\
\eqref{E_s_scaling1}. Thus, we will focus on the second term, which
was zero for the symmetric potential.

We assume again that the matrix element is accumulated at short
distances, where the molecular potential can be expanded in spherical
harmonics~\cite{zeldovich:1977}. The second term of this expansion can
be written as $(\vec{a}\cdot\pos)\hat{V}_\mathrm{a}(\abs{\pos})$,
where $\vec{a}$ is some constant polar vector. In this approximation
we get:
\begin{equation}
 \label{eq: commut3}
 \brackets{\hat{\mathbf{k}},\hat{V}(\abs{\pos})}_-
 = -\imath(\vec{\pauli} \cdot (\pos \times
\vec{a}))\hat{V}_\mathrm{a}(\abs{\vec{r}})
 \,,
 \end{equation}
Substituting this into the second term in eq.\ \eqref{eq: mcommut} we find that:  
\begin{multline}
\Braket{\psi_i|\diraccontra{5}|\psi_i}_\mathrm{a} 
\\
\approx \frac{\alpha^3}{2}
     \Braket{\varphi_i|
     -\imath(\vec{\pauli}\cdot\pos\times\vec{a})
     \hat{V}_\mathrm{a}(\abs{\pos})(\spinmom) 
     |\varphi_i} . 
\end{multline}
Simplifying this further and neglecting the term, which is similar to \eqref{eq: gamma_s3}, we
get: 
\begin{align}\label{eq: gamma5_a}
\Braket{\psi_i|\diraccontra{5}|\psi_i}_\mathrm{a} 
&\approx
             \frac{\alpha^3}{2} \Braket{\varphi_i|
             \vec{a}\cdot\vec{v}_\mathrm{a}|\varphi_i} ,
\\ \label{eq: v_a}
\vec{v}_\mathrm{a} &=              
                                   2\hat{V}_\mathrm{a}(\abs{\pos})\, 
                                   \pos \times \nabla\,.
\end{align}
The orbital pseudovector $\vec{v}_\mathrm{a}$ is $\cal T$-even. The
expected scaling with $\alpha$ is given by eq.\ \eqref{eq: gamma5_a}. 
Scaling with $Z$ for operators \eqref{eq: gamma_s3} and \eqref{eq: v_a} 
should be similar, so we assume: 
 \begin{align}
 \label{E_a_scaling}
 \delta E_{\diraccontra{5},\mathrm{a}}\sim \alpha^3 Z^{2} \,.
 \end{align}

Combining the two terms in eq.\ \eqref{eq: gamma_s-a} together suggests
an estimate for a molecule with two heavy atoms $A$ and $B$:
 \begin{align}
 \label{E_a_scaling2}
 \delta E_{\diraccontra{5}} \approx c_1 \alpha^5 Z_A^{2}Z_B^2
 + c_2 \alpha^3 Z_A^2 + c_3 \alpha^3 Z_B^2 \,.
 \end{align}
The first term is formed on both heavy centers, while the other two terms are formed independently 
in the vicinity of each heavy nucleus. The chiral structure of the molecule is weakly felt locally
\cite{zeldovich:1977,hegstrom:1980}, so we can expect that $\abs{c_{2,3}} \ll \abs{c_1}$.   

In the following we discuss the implications in molecular systems of
the equation derived above for $\Braket{\diraccontra{5}}$ and compare
to results from numerical computations. Hereby, we focus on
scaling with respect to the nuclear charge number and the fine
structure constant. Furthermore, we compare to energy shifts due to
nuclear spin-independent electroweak neutral-current interactions. 

\section{Computational Details}
Quasi-relativistic two-component calculations of \ce{H2X2} and
\ce{H2XO} with X~$=$~O, S, Se, Te, Po and CHBrClF are performed within
the zeroth order regular approximation (ZORA) at the level of complex
generalized Hartree--Fock (cGHF) or Kohn--Sham (cGKS) with a modified
version\cite{wullen:2010,berger:2005,berger:2005a,nahrwold:09,isaev:2012,gaul:2017,gaul:2020}
of the quantum chemistry program package
Turbomole\cite{ahlrichs:1989}. 

For calculations of \ce{H2X2} and \ce{H2XO} compounds a basis set of
25~s, 25~p, 14~d and 11~f uncontracted Gaussian functions with the
exponential coefficients $\alpha_i$ composed as an even-tempered
series by $\alpha_i=a\cdot b^{N-i};~ i=1,\dots,N$ with
$a=\SI{0.02}{\per\bohr\squared}$,
$b=(5/2\times10^{10})^{1/25}\approx 2.606$ and $N=26$ was used
for X $=$ O, S, Se, Te, Po. The largest exponent coefficients of the
s, p, d and f subsets are $5\times10^8~a_0^{-2}$,
$1.91890027\times10^8~a_0^{-2}$, $13300.758~a_0^{-2}$ and
$751.8368350~a_0^{-2}$, respectively. A similar but slightly smaller
basis set (three f functions less) has proven successful in
calculations of parity violating energy shifts in
\ce{H2Po2}\cite{laerdahl:1999,berger:2005}.  The H atom was
represented with the s,p-subset of a decontracted
correlation-consistent basis of quadruple-$\zeta$
quality\cite{dunning:1989}. 

Structure parameters of \ce{H2X2} were chosen as in Refs.
\cite{laerdahl:1999,berger:2005}. For \ce{H2XO} compounds the
equilibrium bond-length of the O--X bond, for X $=$ S, Se, Te, Po was
obtained by full structure optimization at the level of GHF-ZORA. As
convergence criteria an energy change of less than
$10^{-5}~E_\text{h}$ was used.  Bond angles H--O--X and bond distances
H--O of \ce{H2XO} were assumed to be equal to \ce{H2O2} and bond
angles H--X--O and distances H--X were assumed to be equal to
\ce{H2X2}.  Employed structure parameters are summarized in
\prettyref{tab: strucpara}.

Structure parameters, harmonic vibrational wave numbers and normal
coordinates, of
\ce{CHBrClF}, as well as electronic densities and vibrational wave
functions along the C--F stretching mode were employed as described in
Ref.  \cite{berger:2007}.  Electronic densities along other normal
coordinates were
calculated on the level of ZORA-cGHF and ZORA-cGKS with the same basis
set employed in Ref. \cite{berger:2007}.  Properties were calculated
on the levels of ZORA-cGHF and ZORA-cGKS. Used density functionals are
the local density approximation
(LDA)\cite{kohn:1965,vosko:1980,dirac:1930} and the Lee, Yang and Parr
correlation functional (LYP)\cite{lee:1988} with a generalized
gradient exchange functional by Becke (BLYP) \cite{becke:1988} or the
hybrid Becke three parameter exchange functional
(B3LYP)\cite{stephens:1994,vosko:1980,becke:1993,becke:1993a}.

The ZORA-model potential $\tilde{V}(\pos)$ as proposed by van
W\"ullen\cite{wullen:1998} was employed with
additional damping\cite{liu:2002}.

For calculations of two-component wave functions and properties a
finite nucleus was used, described by a normalized spherical Gaussian
nuclear density distribution
$\rho_{\mathrm{nuc},A}(\pos)=\frac{\zeta_A^{3/2}}{\pi^{3/2}}\mathrm{e}^{-\zeta_A\abs{\pos-\pos_A}^2}$,
where $\zeta_A= \frac{3}{2r_{\text{nuc},A}^2}$ and the root mean
square radius $r_{\text{nuc},A}$ of nucleus $A$ was used as suggested
by Visscher and Dyall\cite{visscher:1997}. The mass numbers $A$ were
chosen to correspond to the isotopes $^{1}$H, $^{12}$C, $^{16}$O,
$^{19}$F, $^{32}$S, $^{35}$Cl, $^{79}$Br, $^{80}$Se, $^{130}$Te,
$^{209}$Po. The weak nuclear charges $Q_{\mathrm{W},A}$ of the various
isotopes with charge number $Z_A$ and neutron number $N_A$ were
included as $Q_{\mathrm{W},A} \approx (1-4\sin^2\theta_\mathrm{W})Z_A
- N_A$, where we have used $\sin^2\theta_\mathrm{W} = 0.2319$ as the
numerical value of the Weinberg parameter.

All relativistic expectation values of $\diraccontra{5}$ and
$\hat{H}_\mathrm{ew}$ were calculated with our ZORA property toolbox
approach described in Ref.~\cite{gaul:2020}.

\section{Results}
\subsection{Scaling laws for $\Braket{\diraccontra{5}}$ in molecules}
In order to confirm results of section \ref{molexpec} we performed
quasi-relativistic numerical calculations at the level of ZORA of
$(P)$-enantiomers of \ce{H2X2} compounds with an dihedral angle of
\SI{45}{\degree}, varying X = O, S, Se, Te, Po. These compounds are
established as a common test system for electroweak parity violation
and its scaling behavior with respect to nuclear
charge~\cite{wiesenfeld:1988,laerdahl:1999,stralen:2004,berger:2005,berger:2005a,nahrwold:09,shee:2016}.
In the above scaling law a factor of $\alpha^2Z_B^2$ emerges from
spin-orbit coupling. This factor is in good approximation equal to
$\alpha^2$ in main group element containing molecules with only one
heavy center (see e.g. Refs. \cite{hegstrom:1980}).  Therefore, for a
variation of one heavy X atom while holding the other one fixed as
oxygen atom (\ce{H2XO}) we would expect roughly a scaling of
$\sim\alpha^3Z_A^2$ (corresponding to the second term in
eq.~\eqref{E_a_scaling2}) as the spin-orbit coupling contribution
(corresponding to the first term in eq.~\eqref{E_a_scaling2}) is
suppressed by a factor of $\alpha^2$. 

\begin{figure}
\includegraphics[width=.5\textwidth]{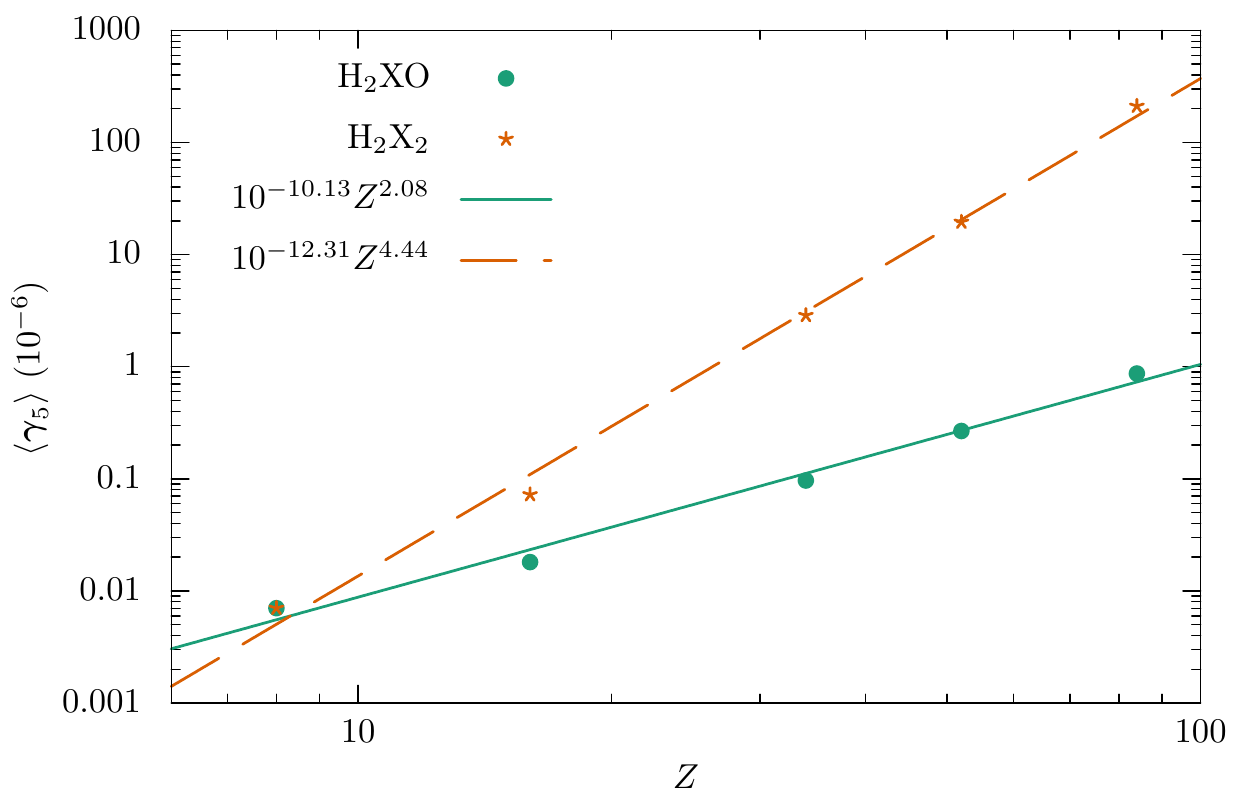}
\caption{Dependence of the expectation value of $\diraccontra{5}$ on the
nuclear charge $Z$ for the $(P)$-enantiomers of \ce{H2X2} and
\ce{H2XO} with X = O,
S, Se, Te, Po at an dihedral angle of \SI{45}{\degree} calculated at the ZORA-cGHF level.}
\label{fig: z_dep}
\end{figure}

The numerical results are summarized in \prettyref{tab: zdep} and
\prettyref{tab: alphadep}.  \prettyref{fig: z_dep} shows a double
logarithmic plot and a linear fit for the determination of the
$Z$-scaling law in ZORA-cGHF calculations. From numerical calculations
of \ce{H2X2} compounds we find a $Z$-scaling with $Z^{4.4}$, which agrees well with the
analytical prediction. Furthermore for
\ce{H2XO} compounds we find a scaling of $Z^{2.1}$, which is in
perfect agreement with the expectations above and shows the missing
spin-orbit coupling contribution as the nuclear charge of oxygen is
close to 1.

In order to test the predicted $\alpha$-dependence the speed of light
was varied in the quasi-relativistic calculations of wave functions
and properties for \ce{H2PoO} and \ce{H2Po2}. The results show the
expected scaling of $\alpha^{5.4}\approx\alpha^5$ for \ce{H2Po2} and a
scaling of $\alpha^{3.6}$ for \ce{H2PoO} showing the weak influence of
spin-orbit coupling in compounds with only one heavy nucleus. The
results are in perfect agreement with the analytical analysis. 

\subsection{Comparison to electroweak electron-nucleon interactions}
Similar considerations, as detailed in the previous section, are known
to hold also for parity non-conserving nuclear spin-independent
electroweak interactions described by Hamiltonian \prettyref{eq:
ewpvoperator} in chiral molecules.  The main difference of this
Hamiltonian to the ones discussed in the theory section is that
$\hat{H}_\mathrm{ew}$ evaluates the expectation value of
$\diraccontra{5}$ at positions inside the nuclei only. To further
compare $\hat{H}_\mathrm{ew}$ with $\diraccontra{5}$ we evaluated the
dependence of the expectation value of both operators on the dihedral
angle in \ce{H2X2} for X $=$ O and Po, and found similar behavior (see
\prettyref{fig: dihedral} and for the explicit data see
the Supplement). It shall be noted, that the sign of
$\hat{H}_\mathrm{ew}$ is inverted in comparison to $\diraccontra{5}$
as $\hat{H}_\mathrm{ew}$ contains in addition the weak charge for
which $Q_\mathrm{W}\approx -N <0$. 

In a recent work \cite{senami:2019}, similar calculations on
$\diraccontra{5}$ in \ce{H2X2} compounds were performed and similar
results were obtained. However, unfortunately, in Ref.
\cite{senami:2019} insufficient basis sets for oxygen were employed
resulting in qualitatively wrong results for the dihedral angle
dependence in \ce{H2O2}. 

The similar dependence on the molecular structure together with the
steep scaling with nuclear charge indicates that contributions at the
nuclear centers dominate also the expectation value of
$\diraccontra{5}$ and, thus, imply that molecular experiments that aim
to test parity violation due to weak interactions can also be used for
searches of parity violating cosmic fields with a comparable
sensitivity. This aspect will be discussed in the following in detail.

\subsection{Limits on cosmic fields from experiments with chiral
molecules}
\subsubsection{Test system and choice of methods}
The expected sensitivity of
experiments with chiral molecules to $\mathcal{P}$-odd cosmic fields
characterized by $b^\mathrm{e}_{0}$
is estimated
from an experiment with \ce{CHBrClF} performed by Daussy \textit{et.
al.}\cite{daussy:1999}, in which a hyperfine component of
the $40_{7,34}\leftarrow 40_{8,33}$ transition
($J'_{K_a',K_c'}\leftarrow J''_{K_a'',K_c''}$) of the C--F stretching
fundamental in enantiomerically enriched samples of the mirror images
$R$-CHBrClF and $S$-CHBrClF was studied. 

Our interest is in a possible splitting of the vibrational resonance
frequency between enantiomers that is caused by cosmic fields
interacting through $\Braket{\diraccontra{5}}$.  For this purpose
frequency shifts in the vibrational spectrum due to electronic
interactions via $\diraccontra{5}$ have to be evaluated.  This test
system, \ce{CHBrClF}, was excessively studied by theory
\cite{quack:2000,quack:2000a,laerdahl:2000a,viglione:2000,quack:2001,schwerdtfeger:2002,schwerdtfeger:2005,berger:2007,thierfelder:2010}
and experiment
\cite{kompanets:1976,bauder:1997,daussy:1999,marrel:2001,ziskind:2002}
and is supposed to be reasonably well understood with respect to
electroweak parity violation.

However, the influence from non-separable anharmonic effects
(multimode effects) on electroweak parity violation in CHBrClF is
largely unexplored. Quack and Stohner studied the deuterated
isotopomer CDBrClF \cite{quack:2003a} with respect to multimode
contributions in a four-dimensional, anharmonically treated subspace
involving the C--F stretch, C--D stretch and the two C--D bending
modes to find an increase of the parity-violating frequency splitting
in the C--F stretch fundamental $\nu_4$ by almost a factor of two ---
depending on the specific model, they obtained up to about 75~\%
relative deviation with respect to the separable anharmonic adiabatic
approximation. Although not directly comparable due to the different
isotope, this at least suggests that pronounced multimode effects can
also exist for $\Braket{\diraccontra{5}}$. 

We have reported major findings and implications for future experiments in a
separate letter \cite{gaul:2020c}, but provide herein more details on
the computational challenges and subsequent analysis.

We estimate the influence of multimode effects within a perturbative
treatment by calculation of derivatives of the property of interest with
respect to all normal coordinates. One-dimensional
and two-dimensional vibrational corrections to a property $O$ for a
single dimensionless reduced normal coordinate $q_r$ are in leading order given
by~\cite{buckingham:1975}:
\begin{align}
O^\mathrm{1D}_{q_r}&\approx \frac{1}{2}\parantheses{v_r+\frac{1}{2}}
\parantheses{ \partiell{^2 O_0}{q_r^2} -
\frac{\phi_{rrr}}{\tilde{\nu}_r}\partiell{
O_0}{q_r}}\\
O^\mathrm{2D}_{q_r}&\approx
-\frac{1}{2}\parantheses{v_r+\frac{1}{2}}\Sum{s\not=r}{}\frac{\phi_{rrs}}{\tilde{\nu}_s}
\partiell{
O_0}{q_s}, 
\end{align}
where $\phi_{rst}$ are the cubic force constants and
$\tilde{\nu}_r$ are the harmonic vibrational wave numbers.

Properties are evaluated along the dimensionless reduced normal coordinate $q_r$
and fitted to a polynomial of degree 4:
\begin{align}
\Braket{\psi_\mathrm{e}|\hat{H}_\mathrm{ew}|\psi_\mathrm{e}}_r&\approx\Sum{k=0}{4}c_{\mathrm{ew},r,k}q_r^k\\
\Braket{\psi_\mathrm{e}|\diraccontra{5}|\psi_\mathrm{e}}_r&\approx\Sum{k=0}{4}c_{\diraccontra{5},r,k}q_i^k\,.
\end{align}

In \prettyref{fig: all_modes} the dependence of
$\Braket{\diraccontra{5}}$ and $\Braket{\hat{H}_\mathrm{ew}}$ on the
normal coordinates for the different methods in the region $q_r=-3,\dots,3$
(for the explicit data see the Supplement). Within this
region the probability density of the first two vibrational states in
the mode $q_4$ is sufficiently decayed (see Fig. 1 of Ref.
\cite{berger:2007}), as can also be expected by considering classical
turning points of a harmonic approximation to the parity-conserving
potential, which are located at $|q_4| = 1$ for the ground vibrational
state of a harmonic oscillator and at $|q_4| = \sqrt{3}$ in the first
vibrationally excited state. The resulting fit parameters
$c_{\diraccontra{5},r,k}$ alongside the explicit values for the
one-dimensional cuts through the hypersurface for all normal coordinates
$q_r$ are reported in the Supplement.

The derivatives of the properties with respect to the normal
coordinate $q_r$ are given by
\begin{align}
\partiell{\Braket{\psi_\mathrm{e}|\diraccontra{5}|\psi_\mathrm{e}}_r}{q_r}
= c_{\diraccontra{5},r,1}\\
\partiell{^2\Braket{\psi_\mathrm{e}|\diraccontra{5}|\psi_\mathrm{e}}_r}{q_r^2}
= 2c_{\diraccontra{5},r,2}\,,
\end{align}
and analogously for $\hat{H}_\mathrm{ew}$. Resulting first and second
derivatives from the fit in \prettyref{fig: all_modes} are listed in
\prettyref{tab: gamma5_grad} and \prettyref{tab: epv_grad}. From these
we see that the C--F stretching mode has a weak influence on
$\Braket{\diraccontra{5}}$ in
comparison to the other modes and, thus, is not an optimal choice for
an experiment. In particular along the deformation normal coordinates $q_9$ (Br--Cl),
$q_8$ (Br--F), $q_3$ (H) and $q_2$ (H) the first derivatives of
$\Braket{\diraccontra{5}}$ are considerably 
larger in magnitude than for $q_4$. The second derivatives with respect to the C-F stretching
coordinate are smaller in absolute value than those first derivatives
mentioned, by about an order of
magnitude (see \prettyref{tab: gamma5_grad} and \prettyref{tab:
epv_grad}). We may assume that anharmonic constants can be roughly of
the order $\phi_{rrr}\sim\mathcal{O}(0.1\tilde{\nu}_r)$ and
$\phi_{rrs}\sim\mathcal{O}(0.01\tilde{\nu}_s)$ or even larger (see e.g.
Ref.~\cite{beil:1996,beil:1997} for some cubic force constants in CDBrClF). 
In total, two-dimensional effects on the C--F stretching mode for
$\Braket{\diraccontra{5}}$ can be on the same order as one-dimensional
vibrational effects. Thus not only the effect of parity violating
interactions on the C-F stretching mode is very weak, but also the
theoretical description is limited by the need of an excellent
description of all modes, which is exceedingly difficult. 

It is important to note, that the use of a different vibrational mode (such
as Br-F ($v_8$) or H ($v_3$) deformation) in CHBrClF can result in
vibrational frequency splittings that are larger by about an order of
magnitude and may reduce error bars considerably. This has to be analyzed
in more detail, however, using anharmonic vibrational force fields.

Due to the resulting large error bars for vibrational corrections for
the C--F stretching mode we do not provide a final value for the
enhancement of $b_0^\mathrm{e}$ in the C--F stretching but rather give
an order of magnitude estimate. 

For this purpose, within the separable anharmonic adiabatic
approximation as described in Ref.~\cite{quack:2000a}, where we follow
for this specific application Ref. \cite{berger:2007} closely,
the vibrationally averaged expectation value for the C--F stretching
mode is evaluated from a series expansion in the vibrational moments
$\Braket{v|q^k|v}$, where $v$ represents the vibrational quantum
number of the $v$th vibrational state. The vibrational wave functions
and corresponding moments were received in Ref. \cite{berger:2007}
from a discrete variable representation on an equidistant grid. The
moments were reported in the supplementary material to Ref.
\cite{berger:2007} and are reused for calculating interactions of
CHBrClF with cosmic fields.

In order to estimate electron correlation effects, for the C--F
stretching mode the vibrationally averaged expectation values where
evaluated at the DFT and HF level, the former with different flavors
of density functionals.  The
results of these methods are compared in \prettyref{tab: methods}.

In previous studies on electroweak parity-violating vibrational
frequency splittings in CHBrClF with density functional approaches
\cite{schwerdtfeger:2005,berger:2007} much reduced variations between
the methods were found for the C--F stretching fundamental as can be
expected by the nearly parallel curves shown in \prettyref{fig:
cfstretch}. In Ref.~\cite{berger:2007} we have observed a spread of
about 20~\%{} from the mean value for the four methods used also in
the present work. The variation amongst the various density
functionals (B3LYP, BLYP and LDA) was below 5~\%{}. In
Ref.~\cite{schwerdtfeger:2005} it was found that B3LYP, BLYP and LDA
estimates deviate by 6~\% or less from the values predicted on the
second order many-body perturbation theory level (MP2), with the
latter method giving also absolute values at the equilibrium structure
that agree well with the corresponding CCSD(T) estimates.
Hartree--Fock based predictions, in contrast, displayed larger
deviations from those of the mentioned density functional
calculations.  Similar trends are observed in the present work (see
\prettyref{tab: methods}), but with more pronounced variations for the
structure dependence of $\Braket{\diraccontra{5}}$ as compared to
$\Braket{\hat{H}_\mathrm{ew}}$: Vibrational splittings vary by about
50~\%{} from the mean value of all four methods, with variations
amongst the density functionals being on the order of 25~\%{} or less
from their mean. Assuming again that the density functionals
outperform the Hartree--Fock approach for this property and give again
similar results as MP2, we are lead to a rough error estimate of about
30~\%{} for the density functionals. Of the different functionals, we
give herein tentative preference to the B3LYP results as i) the
absolute values at the equilibrium structures for electroweak parity
violation were for B3LYP closer to the MP2 and CCSD(T) values
\cite{schwerdtfeger:2005,thierfelder:2010}, ii) the atomic
contributions studied in
Refs.~\cite{schwerdtfeger:2005,thierfelder:2010}, which are
differently weighted by $\Braket{\diraccontra{5}}$ as compared to
$\Braket{\hat{H}_\mathrm{ew}}$, were found to be more consistent with
MP2 and CCSD(T) values and iii) the vibrational splitting on the B3LYP
level is smaller than for the other functionals, which results in more
conservative sensitivity estimates.

\subsubsection{Sensitivity to static cosmic fields}
The expectation values of $\diraccontra{5}$ and splittings between
enantiomers are given in
\prettyref{tab: methods}. As discussed above, we expect multimode
effects of the same size as single-mode effects and at the present
stage are not able to set upper bounds on $b^\mathrm{e}_{0}$ from the
CHBrClF experiment. In Ref.~\cite{gaul:2020c} we rather estimated the 
sensitivity of this
experiment. Assuming B3LYP to give the best performance (see discussion
above) $\Delta_{(R,S)}\Braket{\diraccontra{5}}$ is on the order of
$10^{-10}$ ($\mathcal{O}(10^{-10})$).

The sensitivity of the CHBrClF experiment, performed by Daussy et al. in
1999 \cite{daussy:1999}, to $b^\mathrm{e}_{0}$
was in Ref.~\cite{gaul:2020c} estimated from the experimental upper bound of the parity
violating frequency splitting in the C-F stretching fundamental
$\left|\Delta\nu\right|=\SI[separate-uncertainty=true]{12.7}{\hertz}$\cite{daussy:1999} as:
\begin{align}
\abs{b^\mathrm{e}_{0}}
\lesssim\left|\frac{\SI{12.7}{\hertz}}{\mathcal{O}(10^{-10})}h\right|
\sim\mathcal{O}(10^{-12}\,\si{\giga\electronvolt})
\end{align} 

In comparison to the actual best direct limits on $b^\mathrm{e}_{0}$
from modern atomic experiments, that are
\SI{2e-14}{\giga\electronvolt} from Cs and
\SI{7e-15}{\giga\electronvolt} from Dy\cite{roberts:2014a}, 
the 1999 \ce{CHBrClF} experiment is less sensitive
by about two orders of magnitude \cite{gaul:2020c}. However, it is as sensitive as
atomic experiments with Tl and Yb
($\abs{b^\mathrm{e}_{0}}<\SI{2e-12}{\giga\electronvolt}$, see Ref.
\cite{roberts:2014a}).   

As emphasized in the discussion of multimode effects the sensitivity
of future experiments can be increased by an
order of magnitude, when choosing favorable vibrational transitions.
As we pointed out in Ref. \cite{gaul:2020c}, it was emphasized in Refs.
\cite{ziskind:2002,darquie:2010} that the sensitivity of the experiment
discussed above is improvable by at least two orders of magnitude by experimental
refinement. A choice of a more favorable molecule is expected to lead
to further enhancement by two orders of magnitude. Thus it was estimated
in Ref.~\cite{gaul:2020c} that in future $\mathcal{P}$-violation experiments with chiral
molecules the limits from the 1999 experiment can be improved down to
$10^{-17}\,\si{\giga\electronvolt}$, i.e. an improvement of the actual best
limit by at least two orders of magnitude. This makes experiments with chiral
molecules highly powerful tools to search for Lorentz invariance violation
beyond the Standard Model of particle physics.

The accuracy of the estimate for cosmic field effects in CHBrClF, which was
in this work indirectly inferred by comparison to previous studies on
electroweak parity violation, can in principle be benchmarked by future
explicit calculations with systematically improvable electron correlation
methods and the presently neglected multi-mode contributions can be
accounted for by explicit calculation of anharmonicity constants. As
the main purpose of the present studies was to explore the general potential
of chiral molecules to act as sensitive probes for new physics, more
accurate theoretical estimates specifically for CHBrClF do not seem to be
pressing until new experiments with higher accuracy are performed. Given
the pronounced scaling with nuclear charge that was shown analytically and
confirmed numerically in this paper, the main focus will likely be shifted
to accurate estimates for chiral compounds with heavier elements.
Furthermore, our study showed that care has to be taken by
choice of the vibrational mode, which on the one hand can directly
influence the sensitivity by an order of magnitude and on the other
hand can be crucial for accurate theoretical predictions, which are
essential to provide limits on cosmic fields from experiments.

\section{Conclusion and outlook}
In this paper we have shown that interactions of electrons with the
timelike-component of pseudovector cosmic fields are
strongly pronounced in chiral molecules. Due to the $\mathcal{P}$-odd
contributions of the nuclear potential, that electrons experience in a
chiral molecule, these interactions lead to $\mathcal{P}$-odd
resonance frequency splittings between enantiomers, similar to those
from electroweak parity-violating interactions. We could show
analytically and numerically that these interactions are strongly
enhanced in heavy element containing molecules and are dominated from
contributions that stem from the region near the nucleus. It was
demonstrated that $\mathcal{P}$-odd interactions of electrons with
cosmic fields show similar behavior to interactions due to electroweak
coupling of electrons and nucleons in chiral molecules.  Thus,
knowledge from electroweak quantum chemistry can be employed to find
promising candidate molecules to limit $\mathcal{P}$-odd electronic
coupling to cosmic fields. However, care has to be taken as our
calculations revealed a stronger dependence of $\diraccontra{5}$ on
molecular structure.

We calculated matrix elements of $\mathcal{P}$-odd cosmic field
interactions in \ce{CHBrClF} with quasi-relativistic ab initio
methods, including vibrational corrections, and compared the
results of different DFT functionals. Our calculations of
$\mathcal{P}$-odd effects along the different normal coordinates of CHBrClF
revealed an important role of non-separable
anharmonic effects and showed that the C--F stretching mode in
particular is from this perspective not ideally suited for a measurement of
$\mathcal{P}$-violation due to cosmic fields. Effects on some other modes are expected to be
larger by an order of magnitude. These findings underline the
importance to select not only a favorable molecule, but also to carefully
choose the vibrational transition. However, from our
calculations the sensitivity of a 20 year old
experiment with \ce{CHBrClF} to $\abs{b^\mathrm{e}_0}$ was estimated to be
$\mathcal{O}(10^{-12}\,\si{\giga\electronvolt})$. This sensitivity is
inferior by two orders to the actual best direct measurements drawn
from modern atomic parity violation experiments, but was considered to be
improvable to the order of $\mathcal{O}(10^{-17}\,\si{\giga\electronvolt})$ or
better for static pseudovector fields, which would be an improvement
of the actually best limit on $b^\mathrm{e}_0$ by at least two
orders of magnitude. This demonstrates the specific virtue that studies
on chiral molecules provides in the search for new physics beyond the
standard model.

\begin{acknowledgments}
The authors are grateful to the Mainz Institute for Theoretical
Physics (MITP) for its hospitality and its partial support during the
completion of this work.  The Marburg team gratefully acknowledges
computer time provided by the center for scientific computing (CSC)
Frankfurt and financial support by the Deutsche Forschungsgemeinschaft
via Sonderforschungsbereich 1319 (ELCH) ``Extreme Light for Sensing
and Driving Molecular Chirality''. The work of M.G.K. and T.A.I. was
supported by the Russian Science Foundation (RSF) grant No.
19-12-00157. 
\end{acknowledgments}

%\bibliography{AK.bib}
%\bibliography{./mk}
%
\clearpage
%appendix..tables
\begin{table}
\caption{Molecular structure parameters for compounds of type
\ce{H2X2} and \ce{H2XO} with X = O, S, Se, Te, Po employed in all
calculations. Parameters for \ce{H2X2}, where taken from Refs.
\cite{laerdahl:1999,berger:2005}. O--X bond length determined by full
structure optimization of \ce{H2XO} compounds at the level of
ZORA-cGHF.}
\label{tab: strucpara}
\begin{tabular}{lSSSS}
\toprule
X& {$r(\text{X}-\text{X})$/\AA} & {$r(\text{X}-\text{O})$/\AA}&{$r(\text{X}-\text{H})$/\AA }&
{$\sphericalangle(\text{X}-\text{X}-\text{H})/^\circ$}\\
\midrule
O  &1.490  &1.490  & 0.970&100\\
S  &2.055  &1.627  & 1.352&92 \\
Se &2.480  &1.768  & 1.450&92 \\
Te &2.840  &1.933  & 1.640&92 \\
Po &2.910  &2.057  & 1.740&92 \\
\bottomrule
\end{tabular}
\end{table}
\begin{table}
\begin{threeparttable}
\caption{Electronic expectation value of $\diraccontra{5}$ for
$(P)$-enantiomers of compounds of type \ce{H2XO} and \ce{H2X2} at a
dihedral angle of \SI{45}{\degree} calculated at the level of
ZORA-cGHF.}
\label{tab: zdep}
\begin{tabular}{
S
S[table-number-alignment=center,table-format=3.2,table-figures-exponent=2,round-precision=2,round-mode=places]
S[table-number-alignment=center,table-format=3.2,table-figures-exponent=2,round-precision=2,round-mode=places]
}
\toprule
{\multirow{2}{*}{$Z_{\ce{X}}$}}& \multicolumn{2}{c}{$\Braket{\boldsymbol{\gamma}_5}$}\\
& \ce{H2XO}  &    \ce{H2X2}\\
\midrule
8   & 7.0238779E-09 & 7.0238779e-9  \\ 
16  & 1.8095416E-08 & 7.23453e-8   \\
34  & 9.6569610E-08 & 2.87142e-6   \\
52  & 2.6723137E-07 & 1.9541e-5 \\
84  & 8.6897454E-07 & 2.1140504e-4  \\
\bottomrule
\end{tabular}
\end{threeparttable}
\end{table}
\begin{table}
\begin{threeparttable}
\caption{Electronic expectation value of $\diraccontra{5}$ for
$(P)$-enantiomers of \ce{H2PoO} and \ce{H2Po2} at a
dihedral angle of \SI{45}{\degree} calculated at the level of
ZORA-cGHF for different values of the fine structure constant $\alpha$
including $\alpha_0$ which is $\frac{1}{c}$ in atomic units.}
\label{tab: alphadep}
\begin{tabular}{
c
S[table-number-alignment=center,table-format=3.2,table-figures-exponent=2,round-precision=2,round-mode=places]
S[table-number-alignment=center,table-format=3.2,table-figures-exponent=2,round-precision=2,round-mode=places]
}
\toprule
{\multirow{2}{*}{$\alpha$}} & \multicolumn{2}{c}{$\Braket{\boldsymbol{\gamma}_5}$}\\
& \ce{H2PoO}  &    \ce{H2Po2}\\
\midrule
$\frac{1}{90}$  &  3.1978256E-06 & 1.3761535E-03 \\
{$\alpha_0$}    &  8.6897454E-07 & 2.1140504E-04 \\
$\frac{1}{300}$ &  3.4213971E-08 & 2.1024483E-06 \\
$\frac{1}{400}$ &  1.2287689E-09 & 4.5980319E-07 \\
$\frac{1}{1000}$&  6.6467063E-10 & 4.3341923E-09 \\
\bottomrule
\end{tabular}
\end{threeparttable}
\end{table}
\begin{table}
\begin{threeparttable}
\caption{Molecular expectation value of $\diraccontra{5}$ in 
$(S)$-\ce{CHBrClF} for the vibrational ground state and vibrational first
excited state along the $q_4$ normal coordinate (C-F-stretching mode) at the 
level of
ZORA-cGHF (HF) and ZORA-cGKS with LDA, BLYP and B3LYP functionals
within the separable anharmonic adiabatic approximation.}
\label{tab: methods}
\begin{tabular}{
c
S[table-number-alignment=center,table-format=3.2,table-figures-exponent=2,round-precision=2,round-mode=places]
S[table-number-alignment=center,table-format=3.2,table-figures-exponent=2,round-precision=2,round-mode=places]
S[table-number-alignment=center,table-format=3.2,table-figures-exponent=2,round-precision=2,round-mode=places]
}
\toprule
\multirow{2}{*}{Method} & \multicolumn{3}{c}{$\Braket{\boldsymbol{\gamma}_5}$}\\
\cline{2-4}
&{ $v=0,\,(S)$ } & {$v_4=1,\,(S)$}
&{$v_4=1\leftarrow v=0,\,\Delta_{(R,S)}$}\\
\midrule
HF    & -1.890411663462973e-09 & -1.709861272801248e-09 & 3.611007813234498e-10\\
B3LYP & -8.275996318428361e-09 & -7.905971433672116e-09 & 7.400497695124897e-10\\
BLYP  & -8.273161643811120e-09 & -7.822186458055953e-09 & 9.019503715103358e-10\\
LDA   & -1.206341584667373e-08 & -1.147156377273139e-08 & 1.183704147884697e-09\\
\bottomrule
\end{tabular}
\end{threeparttable}
\end{table}
\begin{table}
\begin{threeparttable}
\caption{One dimensional first and second derivatives of the molecular
expectation value of $\diraccontra{5}$ with respect to the reduced normal
coordinate $q_r$ in $(S)$-\ce{CHBrClF} at the level of
ZORA-cGHF (HF) and ZORA-cGKS with LDA, BLYP and B3LYP functionals.}
\label{tab: gamma5_grad}
\begin{tabular}{
S[table-format=2]
S[table-number-alignment=center,table-format=5.2,round-precision=2,round-mode=places]
S[table-number-alignment=center,table-format=5.2,round-precision=2,round-mode=places]
S[table-number-alignment=center,table-format=5.2,round-precision=2,round-mode=places]
c
S[table-number-alignment=center,table-format=5.2,round-precision=2,round-mode=places]
S[table-number-alignment=center,table-format=5.2,round-precision=2,round-mode=places]
S[table-number-alignment=center,table-format=5.2,round-precision=2,round-mode=places]
}
\toprule
{\multirow{2}{*}{$r$}}
&  \multicolumn{3}{c}{$\partiell{\Braket{\diraccontra{5}}}{q_r}\times10^{9}$} 
&& \multicolumn{3}{c}{$\partiell{^2\Braket{\diraccontra{5}}}{q_r^2}\times10^{9}$} \\
\cline{2-4}\cline{6-8}
      &    {LDA}    &    {B3LYP}   &    {HF}       &&   {LDA}      &  {B3LYP}       &      {HF}         \\ 
\midrule
9 &    14.6739500000 &   11.2114700000 &    5.6262410000 &&    0.2407809000 &    0.1390331000 &   -0.2014038000 \\ 
8 &   -34.0516500000 &  -23.9594900000 &   -3.6074840000 &&   -1.3865550000 &   -0.6235326000 &    0.2898079000 \\ 
7 &    -8.7064520000 &   -6.3546780000 &   -1.0689210000 &&    1.0286310000 &    0.7379298000 &    0.4762792000 \\ 
6 &    -9.1324820000 &   -7.3191210000 &   -0.3857792000 &&    2.4104380000 &    0.9459293000 &   -1.0472080000 \\ 
5 &     8.7457420000 &    6.9595180000 &    2.2615950000 &&   -4.0515970000 &   -2.7867550000 &   -0.7295407000 \\ 
4 &     2.2056210000 &    1.0966660000 &   -0.3135956000 &&    0.2814792000 &    0.2976141000 &    0.4576408000 \\ 
3 &    15.6549600000 &   11.0070800000 &    4.8975730000 &&   -1.9012840000 &   -2.3852860000 &   -4.9424020000 \\ 
2 &     7.8935960000 &   10.4719100000 &   13.5734000000 &&   -1.3729010000 &   -0.6651565000 &    1.2235550000 \\ 
1 &     1.4188330000 &    1.2101120000 &    0.6458217000 &&    0.4617567000 &    0.3759016000 &    0.1733028000 \\ 
\bottomrule
\end{tabular}
\end{threeparttable}
\end{table}
\begin{table}
\begin{threeparttable}
\caption{One dimensional first and second derivatives of the molecular
expectation value of $\hat{H}_\mathrm{ew}$ with respect to the reduced normal
coordinate $q_r$ in $(S)$-\ce{CHBrClF} at the level of
ZORA-cGHF (HF) and ZORA-cGKS with LDA, BLYP and B3LYP functionals.}
\label{tab: epv_grad}
\begin{tabular}{
S[table-format=2]
S[table-number-alignment=center,table-format=5.2,round-precision=2,round-mode=places]
S[table-number-alignment=center,table-format=5.2,round-precision=2,round-mode=places]
S[table-number-alignment=center,table-format=5.2,round-precision=2,round-mode=places]
c
S[table-number-alignment=center,table-format=5.2,round-precision=2,round-mode=places]
S[table-number-alignment=center,table-format=5.2,round-precision=2,round-mode=places]
S[table-number-alignment=center,table-format=5.2,round-precision=2,round-mode=places]
}
\toprule
{\multirow{2}{*}{$r$}}
&
\multicolumn{3}{c}{$\partiell{\Braket{\hat{H}_\mathrm{ew}}}{q_r}\times
10^{18}/E_\mathrm{h}$} 
&&
\multicolumn{3}{c}{$\partiell{^2\Braket{\hat{H}_\mathrm{ew}}}{q_r^2}\times10^{18}/E_\mathrm{h}$} \\
\cline{2-4}\cline{6-8}
      &    {LDA}    &    {B3LYP}   &    {HF}        &&   {LDA}      &  {B3LYP}       &      {HF}         \\
\midrule
9 &    -2.0969000000 &   -1.8961710000 &   -1.4171750000 &&    0.0130313000 &   -0.0100853900 &   -0.0284291300 \\ 
8 &    11.4748700000 &    9.4274270000 &    6.2704560000 &&    0.5241591000 &    0.3759293000 &    0.1289370000 \\ 
7 &     6.9664380000 &    6.2852140000 &    5.1072650000 &&   -0.3698288000 &   -0.2870309000 &   -0.2028923000 \\ 
6 &     3.3677740000 &    2.4532800000 &    1.2359920000 &&   -0.8971686000 &   -0.6187652000 &   -0.1943371000 \\ 
5 &    -2.2412580000 &   -1.7214160000 &   -1.3869460000 &&    1.8740690000 &    1.6054080000 &    1.0513540000 \\ 
4 &     1.9743080000 &    2.0643870000 &    1.9190700000 &&   -0.2741336000 &   -0.3016383000 &   -0.3865502000 \\ 
3 &    -6.6797570000 &   -5.9454760000 &   -5.0355770000 &&    0.5552798000 &    0.4706054000 &    0.3808445000 \\ 
2 &    -6.0084070000 &   -6.5816970000 &   -6.8815690000 &&   -0.4061810000 &   -0.5081706000 &   -0.5727465000 \\ 
1 &     0.4976800000 &    0.3662312000 &    0.0718395700 &&   -0.0876463000 &   -0.0104351500 &    0.0923441300 \\ 
\bottomrule
\end{tabular}

\end{threeparttable}
\end{table}

%appendix..figures
\begin{figure}
\includegraphics[width=.5\textwidth]{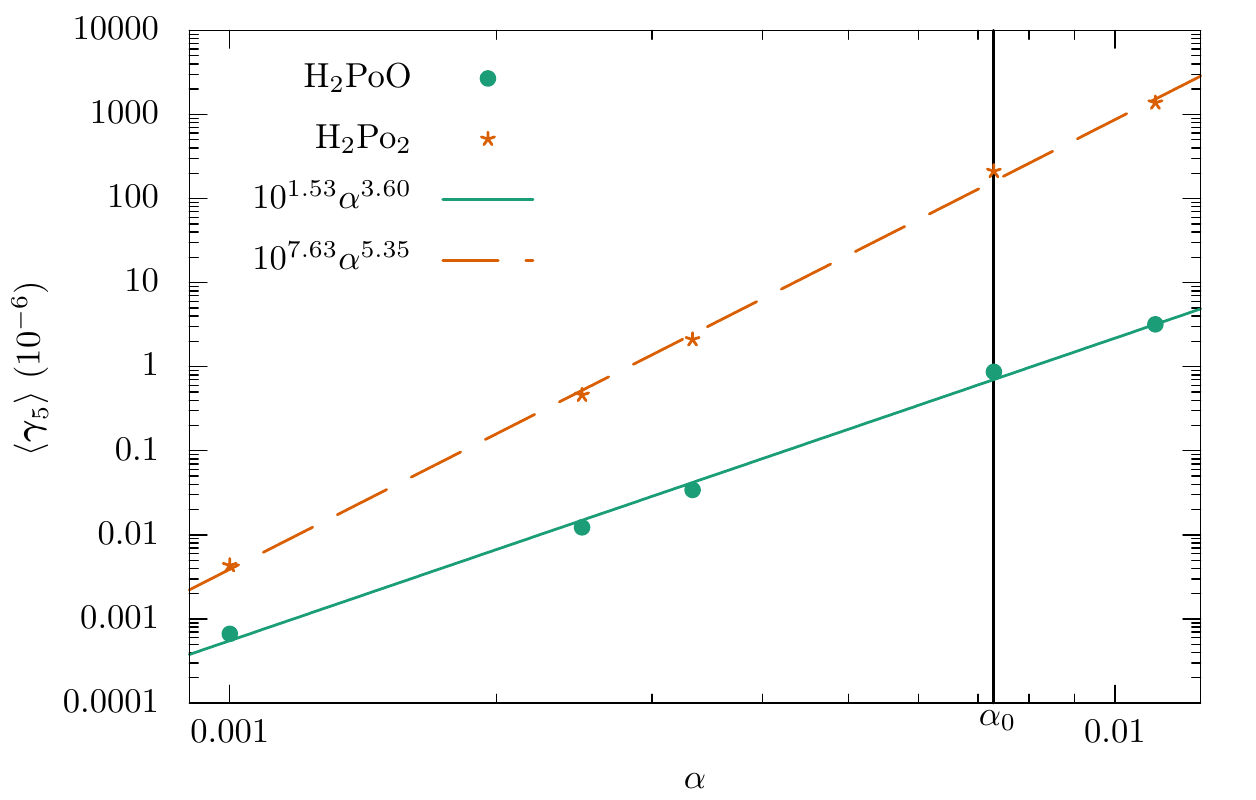}
\caption{Dependence of the expectation value of $\diraccontra{5}$ on
the fine structure constant $\alpha$ for the $(P)$-enantiomers of
\ce{H2Po2} and
\ce{H2PoO} at an dihedral angle of \SI{45}{\degree} calculated at the ZORA-cGHF level.}
\label{fig: alpha_dep}
\end{figure}
\begin{figure}
\includegraphics[width=.5\textwidth]{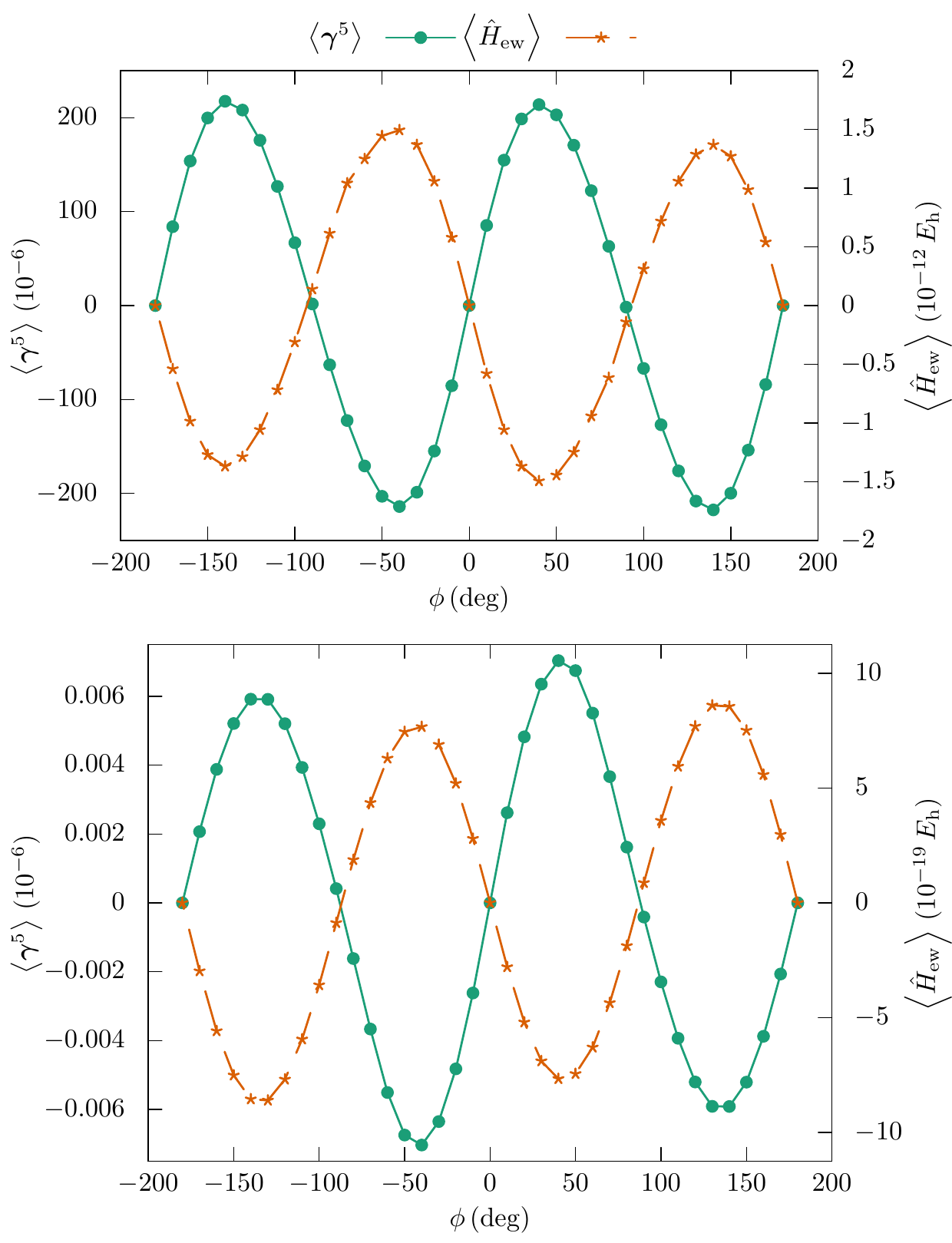}
\caption{Dependence of the expectation value of $\diraccontra{5}$ in
comparison to the expectation value of
$\hat{H}_\mathrm{ew}$ on the
dihedral angle $\phi$ in \ce{H2Po2} (top) and \ce{H2O2} (bottom)
calculated at the ZORA-cGHF level. The results on
$\hat{H}_\mathrm{ew}$ slightly differ from those of Ref.
\cite{berger:2005} due to the use of a different basis set. Straight
lines connecting the computed points are drawn to guide the eye.}
\label{fig: dihedral}
\end{figure}
\begin{figure*}[htp]
\includegraphics[width=\textwidth]{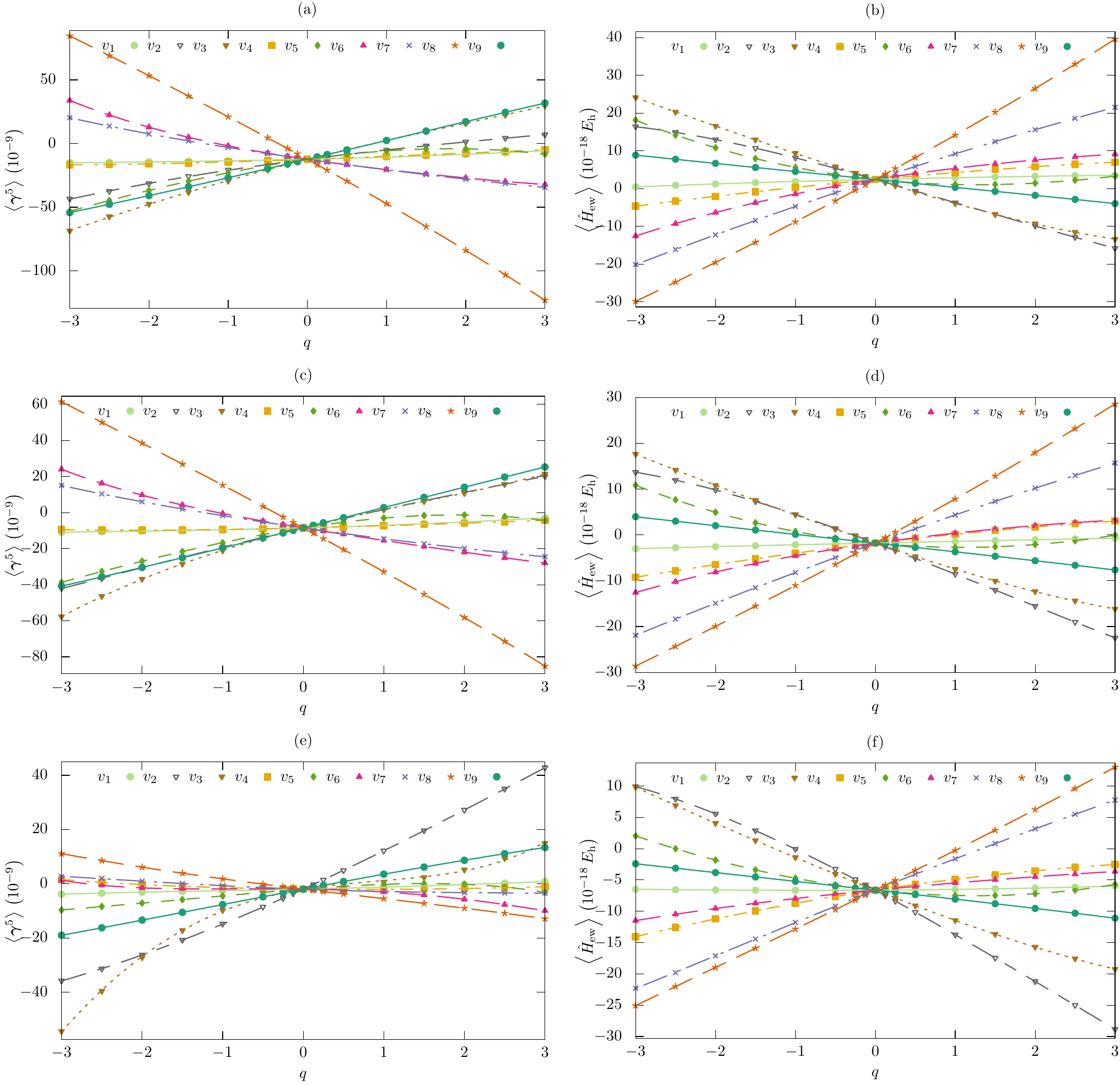}
\caption{Dependence of the expectation value of $\diracco{5}$ (left) 
and $\hat{H}_\mathrm{ew}$ (right) on the nine
normal coordinates in \ce{$(S)$-CHBrClF} computed at the ZORA-cGKS
and ZORA-cGHF level of theory. Data points are fitted to polynomials 
of fourth order (lines).
(a) $\Braket{\diraccontra{5}}$, ZORA-cGKS, LDA; 
(b) $\Braket{\hat{H}_\mathrm{ew}}$, ZORA-cGKS, LDA;
(c) Figure as of Ref.~\cite{gaul:2020c} with values
    corresponding to 
    $\Braket{\diraccontra{5}}$, ZORA-cGKS, B3LYP; 
(d) $\Braket{\hat{H}_\mathrm{ew}}$, ZORA-cGKS, B3LYP;
(e) $\Braket{\diraccontra{5}}$, ZORA-cGHF and
(f) $\Braket{\hat{H}_\mathrm{ew}}$, ZORA-cGHF.
Results for $\hat{H}_\mathrm{ew}$ in the
C--F stretching mode ($v_4$) are a recalculation of those presented in
Ref. \cite{berger:2007} and are thus identical to those.}
\label{fig: all_modes}
\end{figure*}
\begin{figure*}[htp]
\includegraphics[width=\textwidth]{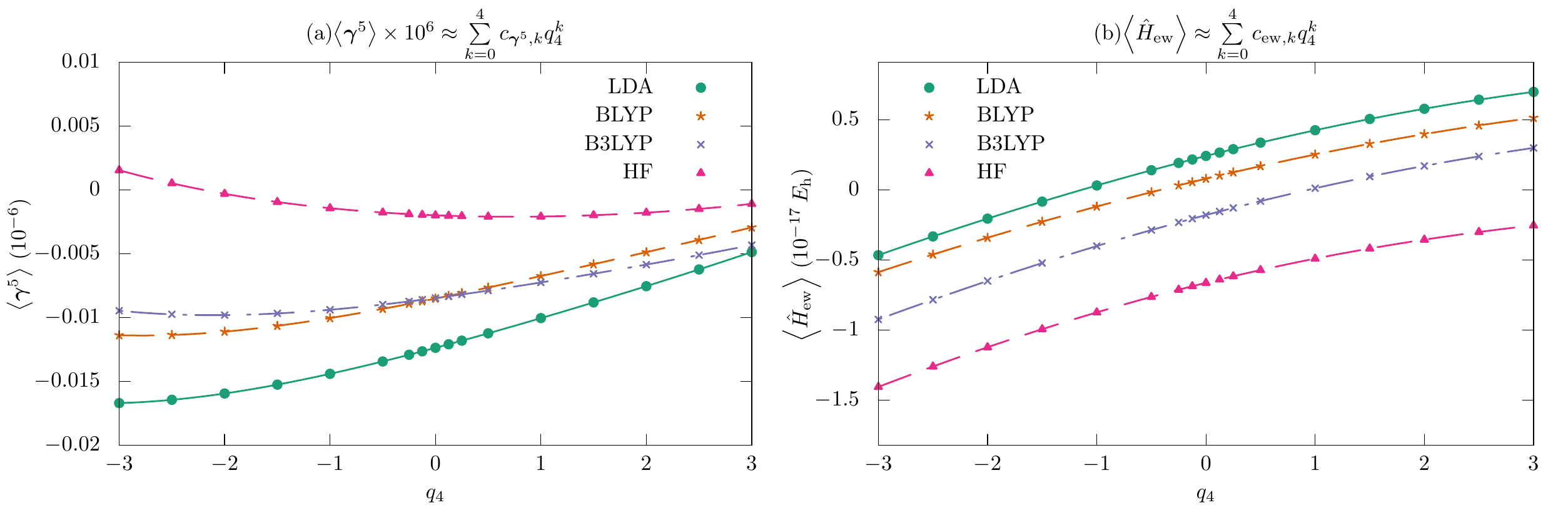}
\caption{Dependence of the expectation value of (a) $\diracco{5}$ and
(b) $\hat{H}_\mathrm{ew}$ on the
C-F stretching normal coordinate $q_4$ in \ce{$(S)$-CHBrClF} computed at the level of ZORA-cGHF and
ZORA-cGKS with different exchange-correlation functionals (points) and polynomial
fits to the $\Braket{\hat{H}_\mathrm{ew}}$ and
$\Braket{\diraccontra{5}}$ to fourth order (lines). The results for
$\hat{H}_\mathrm{ew}$ are a recalculation of those presented in Ref.
\cite{berger:2007} and are thus identical to those.}
\label{fig: cfstretch}
\end{figure*}

\end{document}